\documentclass{article}
\usepackage{graphicx}
\usepackage{pifont}
\usepackage{geometry}
\usepackage[cmex10]{amsmath}
\usepackage{array}
\usepackage{multirow}
\usepackage{bm}
\usepackage{authblk}
\usepackage{mdwmath}
\usepackage{xcolor,soul,framed}
\usepackage{cite}
\usepackage{eqparbox}
\usepackage{url}
\usepackage{amsmath}
\usepackage{multirow}
\usepackage{amssymb}
\usepackage{textcomp}
\usepackage{bm}
\usepackage{dblfloatfix}
\usepackage{svg}
\usepackage{anyfontsize}
\usepackage{tikz}
\usepackage{hyperref}

\definecolor{darkgreen}{RGB}{0,102,51}

\newgeometry{hmargin={20mm,20mm}}   %

\begin{document}

\title{Low voltage user phase reconfiguration as a planning problem 

\let\thefootnote\relax\footnote{
Sari Kerckhove is the corresponding author for this work (e-mail: \texttt{sari.kerckhove@kuleuven.be}). Marta Vanin was affiliated with KU Leuven, ESAT-Electa, Leuven, at the start of this work. She is currently affiliated with KU Leuven, ESAT-Electa, Ghent.
}

}

\author[1,2,3]{Sari Kerckhove}
\author[3,4]{Marta Vanin}
\author[2,3]{Reinhilde D'hulst}
\author[1,3]{Dirk Van Hertem}

\affil[1]{{ESAT - Electa, KU Leuven, Leuven, Belgium}}
\affil[2]{{VITO, Genk, Belgium}}
\affil[3]{{EnergyVille, Genk, Belgium}}
\affil[4]{{ESAT - Electa Ghent, KU Leuven, Ghent, Belgium}}

\date{}

\maketitle

\newcommand{\branches}{$\mathcal{E}$}
\newcommand{\branchesm}{\mathcal{E}}
\newcommand{\branchesi}{$\mathcal{E}_i$}
\newcommand{\branchesim}{\mathcal{E}_i}
\newcommand{\buses}{$\mathcal{N}$}
\newcommand{\busesm}{\mathcal{N}}
\newcommand{\conds}{$\phi$}
\newcommand{\condsm}{\phi}
\newcommand{\condsij}{$\phi_{ij}$}
\newcommand{\condsijm}{\phi_{ij}}
\newcommand{\gens}{$\mathcal{G}$}
\newcommand{\gensm}{\mathcal{G}}
\newcommand{\gensi}{$\mathcal{G}_i$}
\newcommand{\gensim}{\mathcal{G}_i}
\newcommand{\loads}{$\mathcal{L}$}
\newcommand{\loadsm}{\mathcal{L}}
\newcommand{\loadsi}{$\mathcal{L}_i$}
\newcommand{\loadsim}{\mathcal{L}_i}
\newcommand{\meas}{$\mathcal{M}$}
\newcommand{\measm}{\mathcal{M}}
\newcommand{\pseudom}{$\mathcal{P}$}
\newcommand{\mpseudom}{\mathcal{P}}
\newcommand{\rbuses}{$\mathcal{R}$}
\newcommand{\rbusesm}{\mathcal{R}}
\newcommand{\shunts}{$\mathcal{S}$}
\newcommand{\shuntsm}{\mathcal{S}}
\newcommand{\shuntsi}{$\mathcal{S}_i$}
\newcommand{\shuntsim}{\mathcal{S}_i}
\newcommand{\timeseries}{$\mathcal{T}$}
\newcommand{\timeseriesm}{\mathcal{T}}

\newcommand{\phases}{$\Phi$}
\newcommand{\phasesm}{\Phi}

\newcommand{\z}{$\mathbf{z}$}
\newcommand{\h}{$\mathbf{h}$}
\newcommand{\hhm}{$\mathbf{h}_m$}
\newcommand{\errv}{$\boldsymbol{\eta}$}
\newcommand{\vm}{$|U|$}
\newcommand{\va}{$\angle U$}
\newcommand{\vmjp}{$|U_{j,p}|$}
\newcommand{\vajp}{$\angle U_{j,p}$}
\newcommand{\vi}{$U^{\text{im}}$}
\newcommand{\vr}{$U^{\text{re}}$}
\newcommand{\ca}{$\angle I $}
\newcommand{\cax}{$\angle I_c $}
\newcommand{\cix}{$I^{\text{im}}_c$}
\newcommand{\crx}{$I^{\text{re}}_c$}
\newcommand{\cmx}{$|I_c|$}
\newcommand{\px}{$P_c$}
\newcommand{\qx}{$Q_c$}
\newcommand{\w}{$W$}

\newcommand{\cixm}{I^{\text{im}}_c}
\newcommand{\crxm}{I^{\text{re}}_c}
\newcommand{\cmxm}{|I|_c}
\newcommand{\vim}{U^{\text{im}}}
\newcommand{\vrm}{U^{\text{re}}}
\newcommand{\cam}{\angle I}
\newcommand{\caxm}{\angle I_c}
\newcommand{\vmm}{|U|}
\newcommand{\vmmpfi}{|U_i^{\text{pf}}|}
\newcommand{\vmmsei}{|U_i^{\text{se}}|}
\newcommand{\vam}{\angle U}
\newcommand{\pxm}{P_c}
\newcommand{\qxm}{Q_c}
\newcommand{\wm}{W}

\newcommand{\nodevar}{$\dot{x}$}
\newcommand{\edgevar}{$\tilde{x}$}
\newcommand{\nodevarm}{\dot{x}}
\newcommand{\edgevarm}{\tilde{x}}

\newcommand{\branchset}{$\mathcal{E}$}
\newcommand{\revbranchset}{$\mathcal{E}^{\text{R}}$}
\newcommand{\branchsetm}{\mathcal{E}}
\newcommand{\revbranchsetm}{\mathcal{E}^{\text{R}}}
\newcommand{\branchseti}{$\mathcal{E}_i$}
\newcommand{\revbranchseti}{$\mathcal{E}_i^{\text{R}}$}
\newcommand{\branchsetmi}{\mathcal{E}_i}
\newcommand{\revbranchsetmi}{\mathcal{E}_i^{\text{R}}}

\newcommand{\loadset}{$\mathcal{L}$}
\newcommand{\genset}{$\mathcal{G}$}
\newcommand{\loadseti}{$\mathcal{L}_i$}
\newcommand{\genseti}{$\mathcal{G}_i$}
\newcommand{\loadsetm}{\mathcal{L}}
\newcommand{\gensetm}{\mathcal{G}}
\newcommand{\loadsetmi}{\mathcal{L}_i}
\newcommand{\gensetmi}{\mathcal{G}_i}

\newcommand{\singlephaseusersetm}{\mathcal{U}^{1p}}

\newcommand{\N}{\textbf{\textcolor{black}{N}}}

\newcommand{\varspace}{$\mathcal{X}$}
\newcommand{\varspacem}{\mathcal{X}}

\newcommand{\wls}{WLS}
\newcommand{\rwls}{rWLS}
\newcommand{\wlav}{WLAV}
\newcommand{\rwlav}{rWLAV}

\newcommand{\Status}{s}
\newcommand{\Activation}{y}
\newcommand{\Deactivation}{z}
\newcommand{\ApparentPower}{S}
\newcommand{\ApparentPowerVector}{\mathbf{\ApparentPower}}
\newcommand{\ActivePower}{P}
\newcommand{\ActivePowerVector}{\mathbf{\ActivePower}}
\newcommand{\ForecastedActivePower}{\ActivePower^{\text{fx}}}
\newcommand{\GuaranteedActivePower}{\ActivePower^{\text{gtd}}}
\newcommand{\ReactivePower}{Q}
\newcommand{\ReactivePowerVector}{\mathbf{\ReactivePower}}
\newcommand{\ForecastedReactivePower}{\ReactivePower^{\text{fx}}}
\newcommand{\GuaranteedReactivePower}{\ReactivePower^{\text{gtd}}}

\begin{abstract}
Considerable levels of phase imbalance in low voltage (LV) distribution networks imply that grid assets are suboptimally utilized and can cause additional losses, equipment failure and degradation. With the ongoing energy transition, the installation of additional single-phase distributed energy resources may further increase the phase imbalance if no countermeasures are taken.

Phase reconfiguration is a cost-effective solution to reduce imbalance. However, dynamic reconfiguration, through real-time phase swapping of loads using remotely controlled switches, is often impractical because these switches are too costly for widespread installation at LV users. Approaching phase reconfiguration as a planning problem, i.e. static reconfiguration, is an underaddressed but promising alternative. Effective static approaches that allow appropriate imbalance objectives are currently lacking.

This paper presents reliable and expressive static phase reconfiguration methods that grid operators can easily integrate into routine maintenance for effective phase balancing.

We present and compare three static methods, an exact mixed-integer nonlinear formulation (MINLP), a mixed-integer quadratic approximation (MIQP), and a genetic algorithm (GA), each supporting different imbalance objectives. The MIQP approach, despite using proxy objectives, efficiently mitigates the different types of imbalance considered, and outperforms both MINLP and GA in scalability and consistency.
\end{abstract}

\textbf{keywords -}
low voltage distribution grids, phase reconfiguration, planning, imbalance reduction, metaheuristic methods, mathematical optimization.

\section{Introduction}\label{sec:introduction}

\subsection{Problem context}

\begin{table}[b!]
  \caption{Overview of studies considering static phase reconfiguration over multiple time steps.  Abbreviations used: n.k. = not known, n.a. = not applicable, MILP = mixed-integer linear program, DER = distributed energy resource, EV = electric vehicle, PV = photovoltaic }
  \vspace{3mm}
  \renewcommand{\arraystretch}{1.2}
  \setlength{\tabcolsep}{4pt}
  \footnotesize	
  \begin{tabular}{>{\centering\arraybackslash}p{0.05\linewidth}|
                  >{\raggedright\arraybackslash}p{0.10\linewidth}
                  >{\raggedright\arraybackslash}p{0.08\linewidth}
                  >{\raggedright\arraybackslash}p{0.12\linewidth}
                  >{\raggedright\arraybackslash}p{0.14\linewidth}
                  >{\raggedright\arraybackslash}p{0.06\linewidth}
                  >{\raggedright\arraybackslash}p{0.08\linewidth}
                  >{\raggedright\arraybackslash}p{0.1\linewidth}
                  >{\raggedright\arraybackslash}p{0.08\linewidth}}

    Ref.               & Method type& Runtime reported& PF formulation& Objective type& Multiple balance points& Time horizon, and resolution& Testcase            & Limited switching\\
    \hline
    \cite{Abril2014}   & Metaheuristic (other)& No & n.a. (summed loads)& approximate losses, current imbalance& n.a.            & 1 day, hourly& MV, 17 transformers& Yes \\
    \cite{Toma2018}    & Metaheuristic (other)& No & n.a. (summed loads)& current imbalance& Yes& 1 day, hourly& LV, 147 consumers      & No \\
    \cite{Ivanov2019}  & Metaheuristic (GA, other)& No & n.a. (summed loads)& current imbalance& Yes& 1 day, hourly& LV, 68 consumers       & No \\
    \cite{Jimenez2022} & Metaheuristic (other)& No  & n.a. (summed loads)& current imbalance& No& 7 days, 15 min& LV, 250 consumers     & Yes \\
    \cite{benzerga2022optimal} & Metaheuristic (GA) & No & exact& losses, DER curtailment& n.a.            & 1 year, n.k.& LV, 50 DERs           & Yes \\
    \cite{benzerga2022optimal_ev} & Metaheuristic (GA) & No & exact& losses, EV curtailment& n.a.            & n.k. & LV, 10 EVs       & No \\
    \cite{Gangwar2019} & Heuristic (Greedy)      & Yes & exact& losses, current and voltage imbalance& Yes& 30 days, hourly& MV, 123 buses& Yes \\
    \cite{Antic2024}   & Heuristic (Assignment)  & Yes & exact, linearized& PV hosting capacity, voltage imbalance constraint& n.a.            & 1 day, 15 min& LV, 43 consumers& No \\
    \hline 
    \cite{Vassallo2025phase} & MILP              & No & n.a. (summed loads)& current imbalance& No& 1 year, 15  min& LV, 23 consumers       & Yes \\
    \hline 
    This paper          & MINLP                  & Yes & exact& current and voltage imbalance& Yes& 4 days, 15 min& LV, 5 consumers& Yes \\
                        & MIQP                   & Yes & linearized& current and voltage imbalance& Yes& 4 days, 15 min& LV, 100 consumers     & Yes \\
                        & Metaheuristic (GA)     & Yes & exact& current and voltage imbalance& Yes& 4 days, 15 min& LV, 100 consumers     & Yes \\
                        
  \end{tabular}
  \label{tab:related_work}
\end{table}

Phase imbalance is a critical issue in low voltage (LV) distribution networks (DNs) as it results in their underutilization, reducing the hosting capacity of additional low carbon resources~\cite{Wang2013}. As LVDNs are already a bottleneck to electrification and the energy transition~\cite{Antic2024}, reducing phase imbalance is an effective manner to utilize LV DNs more efficiently, increasing their (hosting) capacity while deferring or reducing costly reinforcement actions. 
Imbalanced operation can lead to congestion and voltage issues, as well as higher power losses, thereby raising overall operating costs \cite{Liu2021}. 
Furthermore, at the consumer level, equipment failures may occur; for example, induction motors are susceptible to damage or derating under imbalanced conditions \cite{Soltani2018, Girigoudar2023integration}.

Phase reconfiguration (PR) is a relatively inexpensive and effective way to reduce imbalance \cite{Jimenez2022}. This approach involves physically changing the consumer phase connectivity to achieve a more even distribution of load among the three phases. Other methods to reduce imbalance include power regulation approaches, such as methods using static Var compensators (SVCs) \cite{Zeng2019, Liu2020}, controllable photovoltaic-inverters \cite{ Weckx2013, Girigoudar2020, Girigoudar2023integration}, or using demand-response strategies with (electric vehicle) batteries and home appliances \cite{Zeng2019, Fu2020, Chen2020, tiwari2025incentive}. Network reconfiguration, which involves reconnecting sections of cables to different feeders, is another less direct approach  to reduce imbalance \cite{Siti2007}. In this paper, we will focus solely on phase reconfiguration.

There are two types of phase reconfiguration: \textit{dynamic} real-time approaches, and \textit{static} planning approaches.
The former utilizes remotely controlled phase-switching devices to allow for different phase configurations at each time step. A considerable amount of literature has addressed this \textit{dynamic} problem for both medium voltage (MV) and LV DNs. Mathematical optimization-based approaches to the \textit{dynamic} problem are proposed in \cite{Liu2020, Antic2024, Liu2021Sensitivity, Geng2018, Liu2021, Cui2024}, whereas (meta)heuristic approaches are favored in \cite{Ding2018, Fu2020, Gray2016, Shahnia2014, Soltani2017, Soltani2018, cleenwerck2023smart}. In \cite{Kharrazi2022}, a decentralized approach based on local voltage measurements is proposed. 
Ref. \cite{Liu2021Sensitivity} and \cite{cleenwerck2023smart} propose solutions in case of missing load- or network-data respectively, using sensitivity coefficients and regression to avoid the need for explicit power flow calculations. 
Typically, \textit{dynamic} PR finds the optimal switching actions for one individual time step at a time. If time series data are used as input, each time step is treated independently, i.e., it is not a multi-period problem. Nevertheless, a minority of \textit{dynamic} PR methods are multi-period and consider a limited time horizon of a couple of hours to a day in order to limit the switching frequency \cite{Liu2020, Cui2024, Geng2018}.

A significant obstacle in adopting the \textit{dynamic} approach is that, especially for LV DNs, remotely controllable switches are not available in most grids today and the investments to purchase, support and install these devices are very costly, especially if they require central coordination (factoring in bandwidth and communication costs) \cite{Cui2024}. Additionally, \cite{Jemena} suggests that sudden and repeated phase-switching can negatively impact residential appliances. 

Because of these reasons, it is essential to study the potential of \textit{static} planning approaches to phase-reconfiguration. These require no additional investment costs as the phase connections are changed manually during feeder maintenance. The main challenge with \textit{static} reconfiguration is that, in contrast to the \textit{dynamic} problem, the proposed phase configuration needs to perform well over long time horizons, ranging from multiple months to a few seasons, corresponding to a realistic maintenance frequency. Therefore, projections of demand and generation (as opposed to near-real-time measurements) need to be considered in the optimization. Although some studies have proposed manual (and therefore \textit{static}) PR methods based on a single reputedly representative timestep \cite{Hooshmand2012, Weckx2013, Schweickardt2016, Rios2019, Lin2020, Kay2022, Pereira2021}, these will fall short for long time horizons. A multi-period perspective is essential to account for the diverse load scenarios that may cause imbalance (e.g., peak load versus peak photovoltaic generation). 

Despite its importance, the multi-period \textit{static} reconfiguration problem has not been extensively studied \footnote{While the reason for this limited attention is not entirely clear, it may stem from the problem being explored mainly from a computer science rather than power engineering perspective, together with the belief that automatic switching technologies would soon become widespread in distribution systems, which has not happened in reality.}. The following subsection provides an overview of the (limited) literature on multi-period \textit{static} reconfiguration in DNs and identifies existing gaps.

\subsection{Related work on static phase reconfiguration}

Table \ref{tab:related_work} summarizes recent studies on multi-period \textit{static} phase reconfiguration \cite{Abril2014, Toma2018, Ivanov2019, Jimenez2022, Antic2024, Vassallo2025phase, benzerga2022optimal, benzerga2022optimal_ev, Gangwar2019}.
These studies all exhibit at least one of the following two shortcomings.

Firstly, the majority of the studies in Table \ref{tab:related_work} 
employ (meta)heuristic methods. While these nature-inspired methods provide flexibility regarding nonlinear objectives, they offer no guarantees of finding global optima. For $\text{ill-conditioned problems}$, challenges include: 1) convergence at different local minima with varying fitness values, due to the stochastic nature of metaheuristic methods, and 2) the need for careful hyperparameter tuning and initialization.
These challenges are not addressed in the studies listed in Table \ref{tab:related_work}. Typically, the choice of hyperparameters is not motivated, and the variability of solutions is not reported. For example, \cite{Toma2018} reports only the best solution from 60 runs. 

Secondly, the majority of the methods replace power flow (PF) calculations by summing downstream consumer demands per phase. This reduces accuracy and only allows to minimize current imbalance, leaving more complex objectives such as network losses\footnote{Still, in cases of (meta)heuristic optimization, it should be mentioned that losses can be approximated as done, e.g., in \cite{Abril2014}.} or voltage imbalance incomputable. This limitation can be problematic, since optimal current imbalance reduction may not yield optimal voltage imbalance reduction. In fact, \cite{Girigoudar2020} shows that even closely related imbalance metrics can diverge near optimality, highlighting the need for more expressive \textit{static} multi-period PR methods that enable the accurate computation of a wider range of imbalance metrics.

Dynamic (single-timestep) PR methods do use more expressive mathematical programs and can serve as inspiration for developing a \textit{static} multi-period counterpart: Antic et al. \cite{Antic2024} propose a MINLP based on an exact voltage-current OPF; Liu et al. \cite{Liu2021} offer an iterative method reducing the MINLP to a convex problem; and Liu \cite{Liu2020} and Cui \cite{Cui2024} employ McCormick-envelope linearizations for accurate yet runtime-intensive models. Linearizations using the LinDist3Flow (LD3F) approximation provide a faster alternative that nonetheless works well for voltage-focused objectives, capturing voltage drops accurately despite ignoring line losses \cite{Antic2024, Vanin2020comparison}. While adopting a multi-period formulation increases computational time, this can be acceptable in a planning context where real-time performance is not required. Nonetheless, a careful trade-off between speed and accuracy must be considered, and the scalability with respect to the time horizon should be assessed, a factor that has been largely overlooked in the studies presented in Table \ref{tab:related_work}.

Beyond the discussion of solution method and PF formulation, Table \ref{tab:related_work} further details several properties of practical importance to distribution system operators that should be integrated into real-world phase reconfiguration approaches. These properties include the presence of 'balance points', meaning that balance is optimized at multiple locations along the feeder rather than exclusively at the feeder head, an important aspect when consumer-side balance is required. Additionally, the table indicates whether limits on the maximum number of switched users are considered to reduce reconfiguration costs and time.

\subsection{Contributions}
While \textit{static} phase reconfiguration may be the most cost-effective approach to reduce phase imbalance in LV DNs, it remains insufficiently explored. As shown, important gaps must be addressed to develop viable \textit{static} phase reconfiguration methods. This work presents and compares three novel methods for \textit{static} phase reconfiguration. Two of these are mathematical programming approaches: an exact MINLP formulation, and an MIQP approximation using the LD3F PF formulation, while the third approach is metaheuristic, a genetic algorithm (GA). 

The main contributions of this paper are as follows:
\begin{enumerate}
\item Three novel methods are presented, each allowing a broader range of imbalance objectives, including voltage imbalance, than the models used in previous studies.

\item The used voltage- and power- imbalance objectives are based on the imbalance metrics proposed in the IEEE and the National Electrical Manufacturers Association (NEMA) guidelines. Due to their nonlinearity, two novel proxy imbalance metrics suitable for the MIQP formulation are proposed and their effectiveness is demonstrated.

\item A comparative analysis of the three methods is performed, evaluating the quality of solutions, computational cost, and scalability of the methods with respect to feeder sizes and the considered time horizon. Additionally, the variability in GA solutions, arising from the inherent stochasticity of metaheuristics, which previous studies failed to account for, is assessed.

\end{enumerate}

The structure of the rest of the paper is as follows: in Section \ref{sec:mathematical_model}, the phase reconfiguration problem is mathematically formulated, detailing variable spaces and approximations for MINLP, MIQP, and GA, and introducing the proposed proxy imbalance metrics. Section \ref{sec:ga_implementation} discusses the implementation of the GA. In Section\ref{sec:results} results are presented and analyzed. Finally, Section \ref{sec:conclusion} summarizes key findings and conclusions.

\section{Mathematical Models}\label{sec:mathematical_model}

Let \buses \ and \branches \ be the set of a feeder's nodes (buses) and directed edges (branches), respectively (with $\revbranchsetm$ the reverse set), and let $\phi \in \Phi = \{ 1, 2, 3 \}$ be the three phases of the system. There is a single reference bus $r \in \busesm$ located at the feeder transformer. Let $\singlephaseusersetm$ be the set of single-phase users, and let a unique mapping exist between every user $u \in \singlephaseusersetm$ and the bus $i \in \busesm$ that it is connected to, such that we can use both $u$ and $i$ to index users and their buses. Let $\mathbf{x}$ be the continuous power flow variables that belong to the formulation's variable space (this space depends on the method used as detailed in the following subsections). $\bm{\delta}$ are binary variables that indicate phase connection, e.g., $\bm{\delta}_i = [0, 0, 1]$ means that user $i$ is connected to phase $3$. Bold mathematical symbols and letters represent vectors, if lowercase, and matrices, if uppercase. 

\subsection{Phase reconfiguration}\label{subsec:summaryPR}
The PR problem can be summarized as:

\begin{align}
    \text{minimize} \quad & \mathcal{I}(\mathbf{x}) \label{eq:objective} \\
    \text{subject to:} \quad
    & \mathbf{h}_1(\mathbf{x}) = 0, \label{eq:h_1} \\
    & \mathbf{h}_2(\mathbf{x}, \bm{\delta}) = 0, \label{eq:h_2} \\
    & \mathbf{b}(\bm{\delta}) = 0, \label{eq:b} \\
    & \mathbf{k}(\mathbf{x}) \leq 0, \label{eq:k} \\
    & \mathbf{l}(\bm{\delta}) \leq 0. \label{eq:l}
\end{align}

Where eq.~\eqref{eq:objective} is an ``imbalance objective", chosen from those listed in subsection \ref{subsection:objectives}.
Eq.~\eqref{eq:h_1} represents the power flow equations. Eq.~\eqref{eq:h_2}-\eqref{eq:b}, represent the load injection and phase selection equations respectively, and are thus function of $\bm{\delta}$. Inequality constraints~\eqref{eq:k}-\eqref{eq:l} represent either operational bounds of the system, or a-priori knowledge of the network and user behaviour which helps to reduce the search space. 

The chosen PF formulations for the three different methods are discussed in subsection~\ref{subsec:pf-form}. All equations that refer to phase connectivity (and thus feature binary variables), are further discussed in subsection \ref{subsec:ph-connectivity}. Finally, subsections \ref{subsec:imbal-metrics} and \ref{subsection:objectives} propose appropriate imbalance metrics and objectives, ensuring compatibility with the chosen PF formulation.

\subsection{Power Flow Formulations}\label{subsec:pf-form}
Differences between the PF formulations affect convergence speed, accuracy, and, importantly, which objectives can be used (Subsections \ref{subsec:imbal-metrics} and \ref{subsection:objectives}).

In this subsection, the PF formulations \eqref{eq:h_1} and inequality constraints \eqref{eq:k} for the three methods: MINLP, MIQP, and GA, are discussed.

Because the PR problem is a multiperiod planning problem, all PF equations are considered for each timestep~$t$ in time-horizon $\timeseriesm = \{1, 2,..., T\}$. The continuous variable space $\mathbf{x}$ can be rewritten as $\mathbf{x}~=~\{\mathbf{x}_{\text{pf}},~\mathbf{s}_{i, t}\}$, where $\mathbf{x}_{\text{pf}}$ depends on the PF formulation, while the apparent power injection $\mathbf{s}_{i, t}~=\mathbf{p}_{i, t}+j\mathbf{q}_{i, t}$ $\forall i \in \busesm, t \in \timeseriesm$ is shared for all formulations. \\

The MINLP uses the exact three-wire unbalanced bus injection model in rectangular voltage coordinates from \cite{geth2021real}. Its variable space is summarized as 

$\mathbf{x}_{\text{pf}}^{\text{MINLP}}~=~\{\mathbf{p}_{ij,t},~\mathbf{q}_{ij,t}, \mathbf{u}^\text{re}_{i, t},~\mathbf{u}^\text{im}_{i, t}\in~\mathbb{R}^{3\times 1}\}$, with \(\mathbf{s}_{ij, t}=\mathbf{p}_{ij,t}+j\mathbf{q}_{ij,t}\) the apparent power flow from bus \(i\) to \(j\), and \(\mathbf{u}_{i, t}~=~\mathbf{u}^\text{re}_{i, t}~+~j~\mathbf{u}^\text{im}_{i, t}~\) the voltage at bus $i$.

The PF equations~\eqref{eq:h_1} consist of the power balance equation \ref{eq:minlp_kirchhoff} and generalized Ohm's law~\eqref{eq:minlp_branchflow}
\begin{equation}\label{eq:minlp_kirchhoff}
    \mathbf{s}_{i, t} = \sum_{ij \in \branchsetmi \cup \revbranchsetmi } \mathbf{s}_{ij, t} \; \; \forall i \in \busesm, t \in \timeseriesm,
\end{equation}
\begin{equation}
    \mathbf{s}_{ij,t} = \mathbf{u}_{i,t} \mathbf{Y}_{ij}^*(\mathbf{u}_{j,t}^* - \mathbf{u}_{i,t}^*),\; \; \forall (ij) \in \branchesm \cup \revbranchsetm , t \in \timeseriesm,
    \label{eq:minlp_branchflow}
\end{equation}
with \(\mathbf{Y}_{ij} \in \mathbb{C}^{3\times 3}\) the line admittance matrix and $(\cdot)^*$ the complex conjugate.\\

As mentioned, the MIQP uses the LinDist3Flow PF formulation from \cite{Sankur}. Its variable space is summarized as $\mathbf{x}_{\text{pf}}^{\text{MIQP}}~=~\{\mathbf{p}_{ij,t},~\mathbf{q}_{ij,t},~\bm{\omega}_{i, t}\in~\mathbb{R}^{3\times 1}\}$, where \(\bm{\omega}_{i, t}=|\mathbf{u}_{i,t}|^2\) 
represents the squared voltage magnitude at bus $i$. The PF equations~\eqref{eq:h_1}, consist of the same power balance equation~\eqref{eq:minlp_kirchhoff}, but the generalized Ohm's law becomes:
\begin{equation}\label{eq:LD3Fohm}
    \bm{\omega}_{i, t} = \bm{\omega}_{j, t} + \mathbf{A}_{ij} \cdot \mathbf{p}_{ij, t} + \mathbf{B}_{ij} \cdot \mathbf{q}_{ij, t} \; \; \forall (ij) \in \branchesm , t \in \timeseriesm,
\end{equation}
where
\begin{equation}
\mathbf{A}_{ij} = 2 (\mathfrak{Re}(\bm{\Gamma}) \cdot \mathbf{R}_{ij} + \mathfrak{Im}(\bm{\Gamma}) \cdot \mathbf{X}_{ij}) \; \; \forall (ij) \in \branchesm,
\end{equation}
\begin{equation}
 \mathbf{B}_{ij} = 2 (\mathfrak{Re}(\bm{\Gamma}) \cdot \mathbf{X}_{ij} - \mathfrak{Im}(\bm{\Gamma}) \cdot \mathbf{R}_{ij}) \; \; \forall (ij) \in \branchesm.
\end{equation}
with resistance and reactance matrices $\mathbf{R}_{ij}$ and $\mathbf{X}_{ij}$ $\in \mathbb{R}^{3\times 3}$, and $\bm{\Gamma}$:
\begin{equation}
\bm{\Gamma} = \begin{bmatrix} 
1 & \alpha^2 & \alpha \\
\alpha & 1  & \alpha^2 \\
\alpha^2 & \alpha & 1
\end{bmatrix},
\end{equation}
where $\alpha = e^{-j 2\pi/3 }$.\\

Inequality constraints~\eqref{eq:k} consist of asset overloading (thermal) bounds and voltage bounds. Specifically this includes branch power constraints, bounds on transformer power at the reference bus, voltage magnitude constraints,  voltage angle difference constraints, and fixed voltage magnitude and angle at the reference bus. For the exact constraint formulations in $\mathbf{x}_{\text{pf}}^{\text{MINLP}}$ and  $\mathbf{x}_{\text{pf}}^{\text{MIQP}}$, the reader is referred to \cite{Fobes2020PowerModelsDistribution}. \\

In contrast to the MINLP and MIQP versions of the PR problem, for which the entire set of equations \eqref{eq:objective}-\eqref{eq:l} is directly passed to an off-the-shelf solver, GA decouples the PR problem by iteratively solving a power flow problem (instead of an optimization problem) and separately checking the objective and operational constraints (discussed in more detail in Section \ref{sec:ga_implementation}). This decoupling removes formulation restrictions on the objective and operational constraints, hence defining a variable space $\mathbf{x}_{\text{pf}}^{\text{GA}}$ is only symbolic. In this work, GA uses  
the fixed point iteration current injection method as described in \cite{geth2023implementation} as PF solver.

\subsection{Phase connectivity}\label{subsec:ph-connectivity}

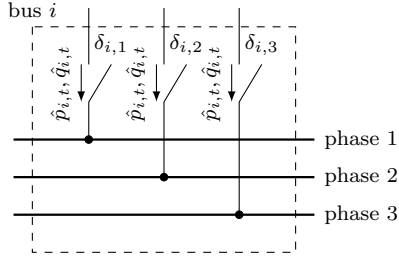
\begin{figure}[t]
    \centering
    \begin{tikzpicture}
        \draw [line width = 0.3mm] (0.0, 0.0) -- (4.0, 0.0) node [right] {\footnotesize phase 1};
        \draw [line width = 0.3mm] (0.0,-0.5) -- (4.0,-0.5) node [right] {\footnotesize phase 2};
        \draw [line width = 0.3mm] (0.0,-1.0) -- (4.0,-1.0) node [right] {\footnotesize phase 3};
        \draw (1.0,1.75) -- (1.0,1.0); 
        \draw [-latex] (0.9,1.0) -- (0.9,0.5) node [pos=0.5, above, rotate=90] {\footnotesize $\hat{p}_{i,t}, \hat{q}_{i,t}$};
        \draw (1.3,1.0) node [above] {\footnotesize $\delta_{i,1}$} -- (1.0,0.5) -- (1.0, 0.0); 
        \draw [fill] (1.0, 0.0) circle (0.5mm);
        \draw (2.0,1.75) -- (2.0,1.0); 
        \draw [-latex] (1.9,1.0) -- (1.9,0.5) node [pos=0.5, above, rotate=90] {\footnotesize $\hat{p}_{i,t}, \hat{q}_{i,t}$};
        \draw (2.3,1.0) node [above] {\footnotesize $\delta_{i,2}$} -- (2.0,0.5) -- (2.0,-0.5); 
        \draw [fill] (2.0,-0.5) circle (0.5mm);
        \draw (3.0,1.75) -- (3.0,1.0); 
        \draw [-latex] (2.9,1.0) -- (2.9,0.5) node [pos=0.5, above, rotate=90] {\footnotesize $\hat{p}_{i,t}, \hat{q}_{i,t}$};
        \draw (3.3,1.0) node [above] {\footnotesize $\delta_{i,3}$} -- (3.0,0.5) -- (3.0,-1.0); 
        \draw [fill] (3.0,-1.0) circle (0.5mm);
        \draw [dashed] (0.25,1.5) node [above] {\footnotesize bus~$i$} rectangle (3.75,-1.5);
    \end{tikzpicture}
    \caption{Model of a single-phase user (figure from \cite{vanin2024phase}).}
    \label{fig:phase-id}
    \vspace{-6mm}
\end{figure}

This section discusses the phase connectivity constraints introduced in 
\eqref{eq:h_2},~\eqref{eq:b}, and~\eqref{eq:l}.  
Fig.~\ref{fig:phase-id} illustrates the concept: single-phase connections are modeled as three-phase, but to each phase $\phi \in \Phi$ a binary variable $\delta_{i,\phi}$ is assigned that determines whether that phase is actually connected ($\delta_{i,\phi} = 1$) or not ($\delta_{i,\phi} =0$). We denote $\bm{\delta}_i = [\delta_{i, \phi}]_{\phi \in \Phi} \in \{0,1\}^{1 \times 3}$. In general, only a subset of the users may be available for phase-reconfiguration $\singlephaseusersetm_{\text{pr}} \subset \singlephaseusersetm$. Hence, $\bm{\delta}_i \; \forall i \in \singlephaseusersetm_\text{pr}$ are variables, while the remaining $\bm{\delta}_i, \; \forall i \in \singlephaseusersetm \setminus \singlephaseusersetm_\text{pr}$ are constants. 
We denote the user load by $\hat{p}_{i, t}\text{~and~}\hat{q}_{i, t}\in\mathbb{R}$, representing active and reactive power at each timestep. Thus, the load injection equation~\eqref{eq:h_2} can be written as:
\begin{equation}
    \hat{p}_{i, t} \cdot \bm{ \delta }_{i} = \mathbf{p}_{i, t} \; \; \forall i \in \singlephaseusersetm, t \in \timeseriesm,
    \label{eq: load-injection p}
\end{equation}
\begin{equation}
    \hat{q}_{i, t} \cdot \bm{ \delta }_{i} = \mathbf{q}_{i, t} \; \; \forall i \in \singlephaseusersetm, t \in \timeseriesm.
        \label{eq: load-injection q}
\end{equation}
Trivially, for buses that are neither connected to  the transformer or a user, there is no injection, hence 
\begin{equation}
    \mathbf{s}_{i, t} = 0, \forall i \in \busesm \setminus (\{r\} \cup \singlephaseusersetm), t \in \timeseriesm.
\end{equation}

The following constraint enforces single-phase connectivity\footnote{Note that only single-phase users are present in the available test cases. Several countries also present three-phase user connections; 
an extension to incorporate three-phase users is straightforward, by, e.g., modeling these as three separate single-phase users as was done in \cite{vanin2024phase}.}: 
\begin{equation}
        \sum_{\phi \in \Phi} \delta_{i, \phi} = 1 \quad \forall i \in \singlephaseusersetm_{\text{pr}},
        \label{eq:deltasum1p}
\end{equation}

Note that~\eqref{eq:deltasum1p} is part of~\eqref{eq:b} and is time-invariant as the phase connectivity is \textit{static} and does not change during the considered time $\timeseriesm$.

Inequality constraints~\eqref{eq:l} consist of limiting the number of switching actions, and, optionally, limiting the number of users per phase. 
The first stems from distribution grid operator practice, and helps to limit the time a maintenance crew would need to perform the reconfiguration. 
A maximum number of actions \({\Delta\delta}^{\text{ max}}\) is enforced:
\begin{equation}
    {\Delta\delta}^{0} = \sum_{i \in \singlephaseusersetm} \max_{\phi \in \Phi} (\delta_{i, \phi}^0 - \delta_{i, \phi}) \leq {\Delta\delta}^{\text{max}},
\label{eq: lim switching actions}
\end{equation}
with $\bm{\delta}^0$ the original phase connectivity of the network.

The second (optional) constraint tightens the feasible space by adding upper and lower bounds to the total number of users connected to each phase, under the assumption that optimal phase assignments lead to an approximately equal number of users per phase (e.g. between $\gamma^{\text{low}}  =  |\singlephaseusersetm| \times 20\% \text{ and } \gamma^{\text{upp}} = |\singlephaseusersetm|  \times 40\%$):
\begin{equation}
    \gamma^{\text{low}} \leq \sum_{i \in \singlephaseusersetm} \delta_{i, \phi} \leq \gamma^{\text{upp}} \; \; \; \forall \phi \in \Phi.
\end{equation}
Note that this assumption is expected to hold in most, but not necessarily \textit{all} scenarios.

\subsection{Imbalance metrics}\label{subsec:imbal-metrics}

Several standard metrics exist to quantify network imbalance. These metrics are defined for a single time step $t$ and locally, i.e.  for a bus $i$ or a branch $ij$. For voltage imbalance, the IEC's Voltage Unbalance Factor (VUF) is defined as the ratio between the negative and positive sequence voltage. Additionally, the NEMA Line Voltage Unbalance Rate (LVUR), and IEEE's Phase Voltage Unbalance Rate (PVUR) are defined as the maximum deviation from the average of respectively the line-to-line, and line-to-ground voltage \cite{Girigoudar2020}. In this work, we adopt the PVUR metric:

\begin{align}
    \text{PVUR}_{i}  = \max_{\phi \in \Phi} \left| 1 - \frac{\big| \mathrm{u}_{i, \phi} \big|}{\left\langle \left| \mathbf{u}_{i} \right| \right\rangle} \right| \times 100 \; \%
\end{align}

with  notation $\langle \left| \mathbf{u} \right| \rangle = \frac{1}{3} \sum_{\phi \in \Phi} \big| \mathrm{u}_{\phi} \big|$. 

NEMA also prescribes the current unbalance rate (I$_\text{U}$) \cite{Jimenez2022}.

\begin{align}
    \text{I}_{\text{U},ij}  = \max_{\phi \in \Phi} \left| 1 - \frac{\big| \mathrm{i}_{ij, \phi} \big|}{\left\langle \left| \mathbf{i}_{ij} \right| \right\rangle} \right| \times 100 \; \%
\end{align}
Similarly, the power unbalance rate can be defined as
\begin{align}
    \text{P}_{\text{U},ij}  = \max_{\phi \in \Phi} \left| 1 - \frac{ \mathrm{p}_{ij, \phi}}{\left\langle \mathbf{p}_{ij} \right\rangle} \right| \times 100 \; \%
\end{align}
As mentioned, the GA method is compatible and scalable with any of these objectives, even when highly nonlinear, as the calculation of the objective is separate from the power-flow calculations. 
In principle, MINLP solvers support nonlinear objective functions, but these may significantly slow down the method. 
For the MIQP approach, none of the metrics above can be directly used, as they cannot be reduced to linear or quadratic expressions in the LinDist3Flow variable space. To address this limitation, two proxy metrics are proposed:  PVUR$^*$ and P$^*_\text{U}$.

PVUR is adjusted by replacing $ |\mathbf{u}_{i}|$ with $\bm{\omega}_i$. Additionally, for the normalization, it is assumed that  $\langle  {\bm{\omega}_{i}} \rangle \approx 1$ (p.u.) to avoid the need for a variable in the denominator:
\begin{align}
    \text{PVUR}^*_{i}  = \max_{\phi \in \Phi} \left| \omega_{i, \phi} - \left\langle \bm{\omega}_{i} \right\rangle \right| \times 100 \; \%
\end{align}

Similarly, the power unbalance rate P$_\text{U}$ is reformulated:

\begin{align}
    \text{P}^*_{\text{U},ij}  = \frac{\sum_{\phi \in \Phi} \left( \mathrm{p}_{ij, \phi} - \mathrm{p}_{ij, (\phi+1) \bmod{3}} \right)^2}{\left( \left\langle \mathbf{p}_{ij} \right\rangle^* \right)^2} \times 100 \; \%
\end{align}

 The quadratic difference is chosen to decrease computation time. The normalizing denominator \({\langle \mathbf{p}_{ij} \rangle}^*\) approximates the average active power flow per phase through summing the downstream demand: 
\begin{equation}
    {\langle \mathbf{p}_{ij} \rangle}^* = \frac{1}{3}\sum_{u \in \singlephaseusersetm \text{, with } ij \in  \mathcal{P}_{ru} } \hat{p}_{u}, 
\end{equation}
where $\mathcal{P}_{ru}$ represents the set of branches making up the path from the reference bus $r$ to user bus $u$. User $u$ is downstream to branch $ij$ if $ij \in \mathcal{P}_{ru}$.

\subsection{Objectives}\label{subsection:objectives}

The location dependent imbalance metrics described in the previous subsection need to be transformed into network-wide imbalance objectives  $\mathcal{I}(\mathbf{x})$  from~\eqref{eq:objective}. For this, multiple ``balance points" can be considered, denoted by $\mathcal{B}^{\busesm} \subset \busesm$ or $\mathcal{B}^{\branchesm} \subset \branchesm$ for bus- or branch-metrics respectively. For voltage unbalance, it is common to consider the balance point with worst imbalance:
\begin{equation}
    \text{PVUR}_t^{(*)} = \max_{i \in \mathcal{B}^{\busesm} } \text{PVUR}_{i, t}^{(*)}.
    \label{eq: PVUR_t}
\end{equation}
Balancing buses are often taken at the users ends (e.g. $\mathcal{B}^{\busesm} = \singlephaseusersetm$). For current/power imbalance, in this paper, the reference bus was picked as the single balance point (which is a common choice in literature) $\mathcal{B}^{\branchesm} = \{r\}$. 
\begin{equation}
    \text{I}_{\text{U}, t} \text{\;or\;}\text{P}_{\text{U},t}^* = \frac{1}{|\mathcal{B}^{\branchesm}|}\sum_{ij \in \mathcal{B}^{\branchesm}} \text{I}_{\text{U}, ij, t} \text{\;or\;}\text{P}_{\text{U},ij, t}^*,
    \label{eq: P_U_t}
\end{equation}
where $|\cdot|$ represents the cardinality. Finally, apart from the location-aspect, there is the time-aspect. As a simple solution, the average over all time steps is taken\footnote{Other papers prefer to use the worst-case time step. However, this might not lead to the best phase assignment in most time steps, while exceeding imbalance limits is permissible for short time. While interesting, investigating time-step/time-series selection methods is out of the scope of this paper.}:
\begin{equation}
    \mathcal{I}(\mathbf{x}) = \frac{1}{|\timeseriesm|} \sum_{t \in \timeseriesm} \mathcal{I}_t (\mathbf{x}).
\end{equation}

\section{Genetic Algorithm implementation}\label{sec:ga_implementation}

This section outlines the implementation of the GA method. Its general workflow is depicted in Figure \ref{fig: ga flowchart}. Problem-specific aspects, including solution encoding and the fitness function design, are covered in the subsections below, followed by a discussion on the choice of selection, mutation, and crossover operators. 

\subsection{Configuration Encoding}

The phase configuration is encoded as a list of integers for each reconfigurable user: \(\mathbf{c} = [\mathrm{c}_u]_{u \in \singlephaseusersetm_\text{pr}}\) with $\mathrm{c}_u\in \Phi$. A mapping between reconfigurable user \(u~\in~\singlephaseusersetm_\text{pr}\) and $i~\in~[1,..., |\singlephaseusersetm_\text{pr}|] $ is defined such that $\mathbf{c}$ can be treated as an ordered list, and indices $u$ and $i$ can be used interchangeably. Note that the phase configuration \(\mathbf{c}\) is equivalent to the binary phase connectivity variables \(\bm{\delta}\):
\begin{equation}
    \mathrm{c}_i = \arg\max_{\phi \in \Phi} \delta_{i, \phi}, \; \text{and} \; \; \bm{\delta}_{i} = [ \mathbf{1}_{\mathrm{c}_i = \phi}]_{\phi \in \Phi}, \; \; \forall i \in \singlephaseusersetm_\text{pr},
\end{equation}
with $\mathbf{1}_{(\cdot)}$ the indicator function.
Integer encoding
 inherently satisfies the single-phase connectivity equation  \eqref{eq:deltasum1p} and is generally a logical choice given the three discrete phases.

\subsection{Fitness Function}

To search the configuration space, the GA is guided by a fitness function $\mathcal{F}(\mathbf{c})$ to minimize:
 \begin{equation}
    \min_{\mathbf{c}} \mathcal{F}(\mathbf{c})
\end{equation}

Given \(\mathbf{c}\) (and therefore $\bm{\delta}$), and the load injection equations, a PF solver can determine \(\mathbf{x}\).
Hence from the full PR-problem \eqref{eq:objective}-\eqref{eq:l},  the equality constraints,~\eqref{eq:h_1}-\eqref{eq:b}, are already satisfied. The remaining equations, namely the imbalance objective~\eqref{eq:objective} $ \mathcal{I}(\mathbf{x})$ and  inequality constraints ~\eqref{eq:k} $(\mathbf{k}(\mathbf{x})\leq 0)$ and ~\eqref{eq:l} $(\mathbf{l}(\mathbf{c})\leq0)$ are taken into account through the fitness function:

\begin{equation}
    \mathcal{F}(\mathbf{c}) = 
    \begin{cases} 
         \text{M}\;\mathcal{I}^0 \quad \text{if not \eqref{eq:l}}, \\ 
        \mathcal{I}(\mathbf{x}) + \text{M}\;\mathcal{I}^0 \; \mathbf{1}_{ \text{[not \eqref{eq:k}]}} \quad \text{otherwise},
    \end{cases}
    \label{eq: fitness function}
\end{equation}

with  $\mathcal{I}^0$  the imbalance objective value for the original phase configuration of the feeder $\mathbf{c}^0 $ ($\Leftrightarrow{}\bm{\delta^0}$) , and M a large number.  From this formula, it becomes clear that the fitness function equals the imbalance objective function $ \mathcal{I}(\mathbf{x})$~\eqref{eq:objective}, unless any of the inequality constraints,~\eqref{eq:k} and~\eqref{eq:l}, are not satisfied. In that case, a large penalty, $\text{M}\;\mathcal{I}^0$, is applied to ensure that any unfeasible configuration will have a fitness that is many times higher than feasible candidates. In this paper, $\text{M}$ is set to 100. To avoid unnecessary PF calculations, the binary constraints, \eqref{eq:l}, are checked first. Only if these are satisfied, the PF solver is used to determine $\mathbf{x}$, and the objective value and operational constraints \eqref{eq:k}  are checked.



\begin{table}[b]
        \caption{Results of MIQP and MINLP with proxy objective PVUR$^*$, and average results of 20 GA-runs with proxy and exact objectives PVUR$^*$ and PVUR. The optimization is done for four days with quarterly hour intervals.}
        \small
    \begin{tabular}{l|>{\raggedright\arraybackslash}p{0.07\linewidth}>{\raggedright\arraybackslash}p{0.07\linewidth}>{\raggedright\arraybackslash}p{0.07\linewidth}|>{\raggedright\arraybackslash}p{0.1\linewidth}>{\raggedright\arraybackslash}p{0.1\linewidth}>{\raggedright\arraybackslash}p{0.1\linewidth}>{\raggedright\arraybackslash}p{0.1\linewidth}>{\raggedright\arraybackslash}p{0.1\linewidth}}
         &   t (s)  &$\#$ threads&$\#$ $f_{calls}$&\textbf{PVUR$^*$ ($ \%$)} &\textbf{PVUR ($ \%$)}& P$_U^*$ ($ \%$)&P$_U$ ($ \%$) &P$_{loss}$ ($\%$)\\
         \hline
         $\bm{\delta}^0$& n.a.&n.a.&n.a.& \textbf{1.86}&\textbf{0.96}& 38.9&24.4 &1.41\\
 MIQP& 3 10$^3$& 10& n.a.&\textbf{1.36 }& \textbf{0.70}& 19.6& 17.1 &1.33\\
 GA - PVUR&  7 10$^3$&10&6 10$^3$& \textbf{1.40}& \textbf{0.72}& 21.0& 17.9&1.34
\\
 GA - PVUR*& 7 10$^3$&10& 6 10$^3$& \textbf{1.39}& \textbf{0.72}& 20.8& 17.8&1.34
\\
 MINLP& n.c.  &n.c.&n.a.&\textbf{n.c.}&\textbf{n.c.}&n.c. &n.c. &n.c.\\\end{tabular}

    \label{tab:opt_results_pvur_approx}
\end{table}

\begin{table}[b]
        \caption{Results of MIQP and MINLP with proxy objective P$_\text{U}^*$, and average results of 20 GA-runs with proxy and exact objectives P$_\text{U}^*$ and P$_\text{U}$. The optimization is done for four days with quarterly hour intervals.}
        \small
    \begin{tabular}{l|>{\raggedright\arraybackslash}p{0.07\linewidth}>{\raggedright\arraybackslash}p{0.07\linewidth}>{\raggedright\arraybackslash}p{0.07\linewidth}|>{\raggedright\arraybackslash}p{0.1\linewidth}>{\raggedright\arraybackslash}p{0.1\linewidth}>{\raggedright\arraybackslash}p{0.1\linewidth}>{\raggedright\arraybackslash}p{0.1\linewidth}>{\raggedright\arraybackslash}p{0.1\linewidth}}
         &   t (s)  &$\#$ threads&$\#$ $f_{calls}$&PVUR$^*$ ($ \%$)&PVUR ($ \%$)& \textbf{P$_U^*$} ($ \%$)&\textbf{P$_U$ ($ \%$)} &P$_{loss}$ ($\%$)\\
         \hline
         $\bm{\delta}^0$& n.a.&n.a.&n.a.& 1.86&0.96& \textbf{38.9}&\textbf{24.4} &1.41\\
 MIQP& 6 10$^3$& 10& n.a.& 1.57& 0.81& \textbf{14.2 }
 & \textbf{14.6} &1.36\\
 GA - P$_U$&  7 10$^3$&10&6 10$^3$& 1.59& 0.82& \textbf{16.2}&\textbf{15.5}&1.36
\\
 GA - P$^*_U$& 7 10$^3$&10& 6 10$^3$& 1.66& 0.85& \textbf{16.1}& \textbf{15.7}&1.37
\\
 MINLP& n.c.  &n.c.&n.a.& n.c.& n.c.&\textbf{n.c.} &\textbf{n.c.} &n.c.\\\end{tabular}

    \label{tab:opt_results_pu_approx}
\end{table}

\subsection{GA architecture and operators}

\begin{figure}
    \centering
    \includegraphics[width=0.37\linewidth]{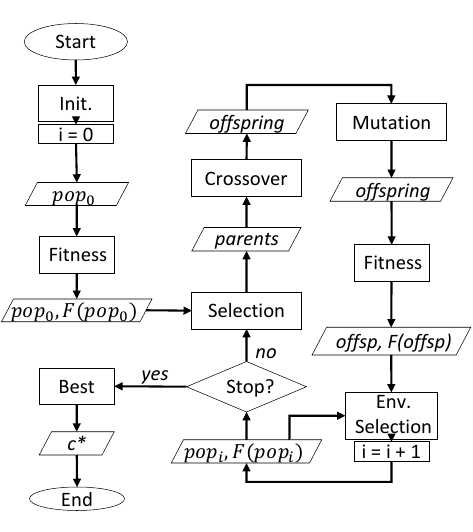}
    \caption{GA's general workflow}
    \label{fig: ga flowchart}
    \vspace{-6mm}
\end{figure}

The GA’s general workflow, following a standard approach, is depicted in Figure \ref{fig: ga flowchart}. First, a starting population ($pop_0$) containing $p$ configurations is initialized (hyperparameter: population size $p$). This initial population always includes the original configuration $\mathbf{c}^0$, with the remaining configurations being randomly initialized. Consecutively, the method iteratively explores the feasible configuration space as follows:
\begin{itemize}
    \item Selection: From the current population ($pop_i$), $p/2$ pairs of parents are selected with \textit{binary tournament selection}. 
    \item Crossover: Pairs of offspring are generated from the parent pairs through \textit{single point crossover} occuring with a crossover probability \(P_c\) (hyperparameter \(P_c \in [0,1]\)).
    \item Mutation: To introduce variability, the offspring are further subjected to \textit{random resetting mutation} for which each element of the parent configuration can be reset to a random phase with mutation probability \(P_m\) (hyperparameter \(P_m\) set to $1/|\singlephaseusersetm_\text{pr}|$ by default).
             
    \item  Environmental selection: Finally, the next generation $pop_{i+1}$ is built from both $pop_i$ and the newly generated offspring through an \textit{elitist replacement} environmental selection operator, passing on the fittest $p$ configurations to the next generation.
\end{itemize}

These are all standard GA operators \cite{kramer2017genetic}. The steps are applied iteratively until the stopping criterion is reached, in this case a maximum number of fitness function calls (hyperparameter $f_{calls}^\text{max}$). Note that total number of generations ($nr_{gen}$ ) can be derived as $\lfloor f_{calls}^\text{max}/p\rfloor$.  The fittest configuration $\textbf{c}^*$ is returned as the solution. In section \ref{subsec:implementation-hyperparam}, the choice of hyperparameters is discussed.

\section{Test cases and Implementation}\label{sec:implementation}

\subsection{Test cases (TC)}

The proposed methods are evaluated on several three-phase three-wire feeders from the ENWL dataset \cite{espinosa2015low}. Load profiles from the NREL dataset \cite{Wilson2022} (108 load profiles from the state of New York, quarter-hour resolution, 2022) are randomly assigned to each consumer. Three test cases are defined:
\begin{itemize}
    \item \textit{TC1 - Methods performance and comparison:} Conducted on a 55-consumer feeder (ENWLs network 1 feeder 1 or the IEEE LV feeder). Optimization is performed for the first four days of January (96 time steps/day). MINLP, MIQP, and GA are tested for imbalance objectives PVUR and P$_\text{U}$ (or proxies), with GA repeated 20 times to account for variability. The MIQP solutions are additionally validated for unseen loads from the full year.
    
    \item \textit{TC2 - Impact of the number of switching actions:} Examines the impact of varying $\Delta{\bm{\delta}^\text{max}}$ using MIQP on the LV test feeder ($|\singlephaseusersetm| = 55$) over a single day (24 time steps, hourly resolution).
    
    \item \textit{TC3 - Scalability assessment:} Evaluates computation times across ENWL feeders of varying sizes (20, 55, 72, 100 consumers) and different optimization horizons (1, 24, 96, 192, 384 time steps). Tested for P$_\text{U}^*$ with GA repeated 5 times per setting; the slowest GA run determines $f_{calls}^\text{max}$ and the reported time.
\end{itemize}

\subsection{Implementation} \label{subsec:implementation-hyperparam}

All methods are implemented in Julia. The MIQP and MINLP optimization methods are build on the \textit{PowerModelsDistribution.jl} package \cite{Fobes2020PowerModelsDistribution} with additional imbalance objectives, binary variables and related constraints. They are solved using Gurobi for MIQP and Juniper \cite{kroger2018juniper} (with Gurobi for the mixed-integer component and Ipopt \cite{wachter2006ipopt} for the nonlinear component) for MINLP. The simulations were run on a 64-bit server with two Intel Xeon Silver 4210 processors @2.2 GHz. 
The number of threads was set to 10 (passed on as a Gurobi parameter).

The GA is implemented as an extension of \textit{Metaheuristics.jl} \cite{Mejia-de-Dios2022metaheuristics}, with power flow calculations performed using \textit{PowerModelsDistribution}'s native solver. GA simulations were executed on a 64-bit server with an Intel Xeon E5-2690 v3 processor @2.6 GHz. For comparability, the GA was also assigned 10 threads for parallelization.

For the GA hyperparameters, population size $p$ was set to $100$, $f^\text{max}_{calls}$ is set to $6\cdot 10^3$ (except for \textit{TC3}, discussed above), crossover probability $P_c$ is $0.7$, while the mutation probability $P_m$ is $1/|\singlephaseusersetm_\text{pr}|$, meaning that, on average, one consumer is rephased per mutation. These choices were partially guided by hyperparameter optimization; however, the results proved inconclusive, as the variability in GA solutions was greater than anticipated.

Finally, key problem parameters are set as follows: user limits per phase range from $\gamma^{\text{low}} = 20\%$ to $\gamma^{\text{upp}} = 40\%$, $\Delta\delta^{\text{max}}$ is set to 5 switching actions (except for dedicated test case \textit{TC2}). All consumers are available for phase reconfiguration, hence $\singlephaseusersetm_{\text{pr}}=\singlephaseusersetm$. The original phase configuration $\bm{\delta}^0$ ($\Leftrightarrow \mathbf{c}^0$) is taken as the connectivity present in the dataset.

\section{Results}\label{sec:results}


\subsection{Methods performance and comparison (TC1)}


Tables \ref{tab:opt_results_pvur_approx}, and \ref{tab:opt_results_pu_approx} present the results of \textit{TC1} for PVUR($^*$) and P$_\text{U}$($^*$)  respectively. For GA, the average results out of 20 runs are reported.

MINLP did not converge for any of the objectives within the given time-limit of 24 hours. MIQP and GA did converge and both methods achieved significant imbalance reduction: around 27\%, 27\%, 63\% and 40\% for PVUR$^*$, PVUR, P$_\text{U}^*$, P$_\text{U}$.  Not only the imbalance objective under optimization, but also the other imbalance objectives are simultaneously reduced. For example, MIQP, when optimized for PVUR$^*$, achieved 27\%, 27\%, 50\%, 30\%, and 6\% reduction in PVUR$^*$, PVUR, P$_\text{U}^*$, P$_\text{U}$, and power losses P$_{loss}$ (Table \ref{tab:opt_results_pvur_approx}). Similarly, optimizing for P$_\text{U}^*$ resulted in 16\%, 16\%, 63\%, 40\%, 4\% reduction in  PVUR$^*$, PVUR, P$_\text{U}^*$, P$_\text{U}$, and P$_{loss}$(Table \ref{tab:opt_results_pu_approx}). Despite this simultaneous reduction in all imbalance metrics for each objective, the choice of objective remains important since the amount of reduction varies: When PVUR* is the objective, PVUR and PVUR$^*$ are reduced more than P$_\text{U}$ and P$_\text{U}^*$. The same holds the other way around. This result further illustrates that the proxy objectives are successful in reducing their non-convex counterparts.

When optimizing for PVUR$^*$ (results from Table \ref{tab:opt_results_pvur_approx}), the MIQP method reduces PVUR$^*$ imbalance with 27\%, 
while GA achieves less reduction: 25\% for the average run (27\% and 22\% for the best and worst runs out of 20, see Figure \ref{fig:GAboxplots}). Moreover, evaluated for PVUR imbalance,  MIQP optimized for proxy objective PVUR$^*$ achieves better imbalance reduction than the average GA-run optimized for exact objective PVUR, namely 27\% 
compared to 25\% on average. 
Similarly, when optimizing for P$_\text{U}^*$ (results from Table \ref{tab:opt_results_pu_approx}) \, the MIQP method reduces P$_\text{U}^*$ imbalance more than GA does (63\% compared to an average of 59\%). Again, evaluated for P$_\text{U}$, MIQP optimized for the proxy  P$_\text{U}^*$ achieves better results than the average GA-run optimized for the actual objective P$_\text{U}$ (40\%  compared to 36\% average). 

Finally, the computation times are considered.  Note that both GA and MIQP are parallelizable methods due to parallel fitness computation for GA on the one hand (additionally, the PF computation could be further parallelized for each time step), and the branch and bound algorithm for MIQP on the other hand. Both methods were given 10 parallel processes to make a fair comparison. For GA, the computation time is strongly connected to the maximum number of fitness function calls, hyperparameter $f_{calls}^\text{max}$. This hyperparameter is set to $6 \cdot 10^3$ (for which 90\% of the GA runs converge. Figure \ref{fig:GA convergence} illustrates this for the power unbalance objective P$_\text{U}$).  With little variation, GA takes approximately 110 min for each run and all objectives. MIQP takes 50 min when solving for PVUR$^*$, and 100 min when solving for P$_\text{U}^*$ so there is a significant time difference between different objectives but a better overall performance than GA.

In conclusion, despite the prevalence of metaheuristic phase-reconfiguration methods proposed in literature, this study finds that the GA method is slower and has lower quality solutions than the MIQP method using the LinDist3Flow approximate power flow and  proxy objectives, even though it has the capability to accommodate non-convex objective functions.

\subsection{GA performance variability (TC1)}

\begin{figure}[t!]
    \centering
    \includegraphics[width=0.5 \linewidth]{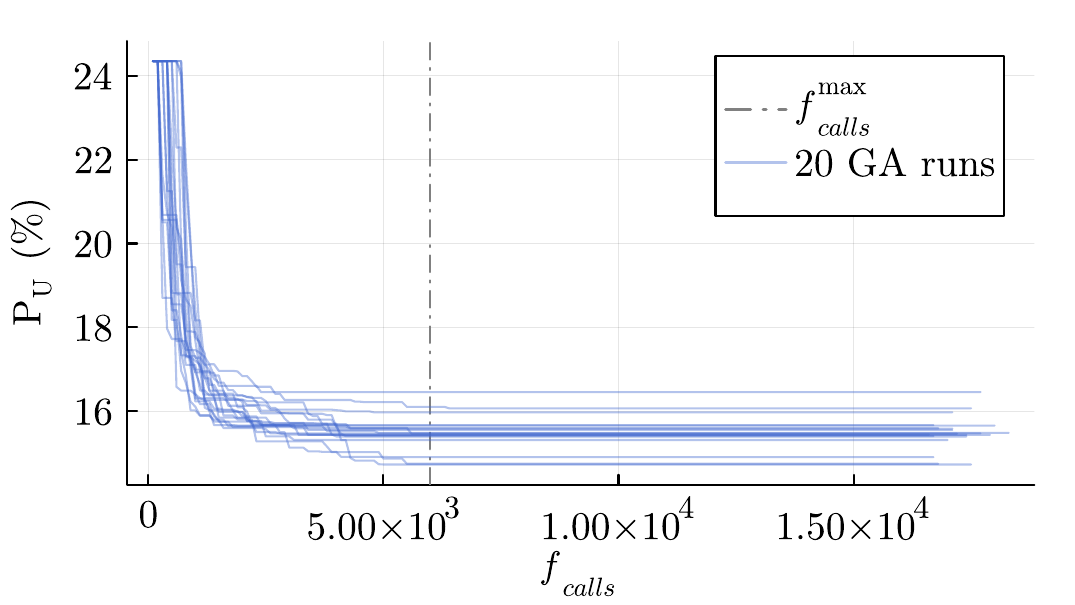}
    \caption{Convergence of GA with objective P$_\text{U}$ in function of the number of fitness function calls for each of the 20 runs. $f_{calls}^\text{max}$ is determined to be $6 \cdot 10^3$. }
    \label{fig:GA convergence}
    \vspace{-3mm}
\end{figure}

\begin{figure}
    \centering
    \includegraphics[width=0.5 \linewidth]{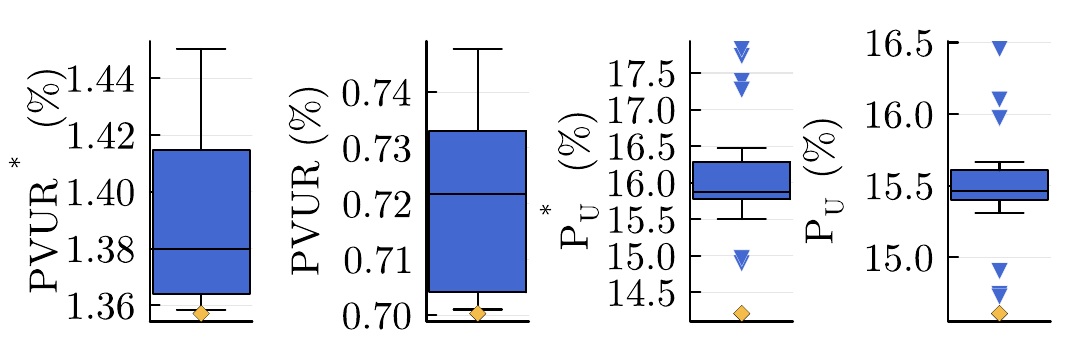}
    \caption{Distribution of imbalance/fitness for each objective for 20 GA-results. The MIQP result is represented by the yellow diamond mark at the bottom of the boxplots.}
    \label{fig:GAboxplots}
    \vspace{-3mm}
\end{figure}

As mentioned, GA was run 20 times for each objective. 
Figure \ref{fig:GA convergence} illustrates the fitness evolution of each run for objective P$_\text{U}$. A significant variation in fitness values is observed even after convergence. Similar large variations are found for the other objectives, as shown in the box-plots in Figure \ref{fig:GAboxplots}, indicating that GA’s tendency to get stuck in local minima is a particular issue for the PR problem. For each objective, the corresponding MIQP result, highlighted in yellow, performs invariably better or equally good as the best GA run.

\begin{figure*}[t!]
    \centering
    {\includegraphics[width=0.475\textwidth]{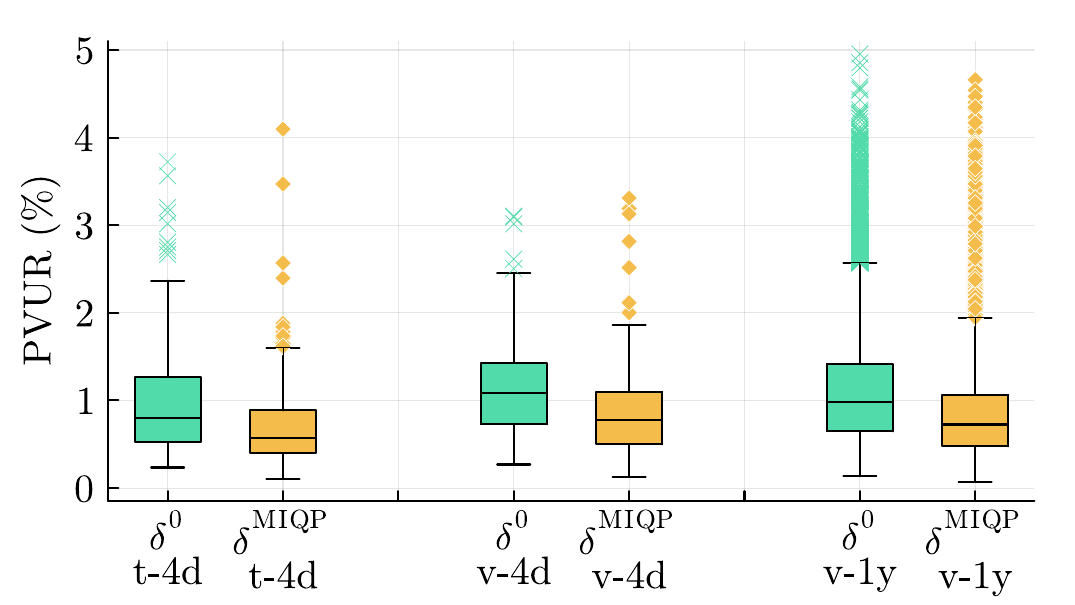}\label{fig:a}}
    \qquad
    {\includegraphics[width=0.475\textwidth]{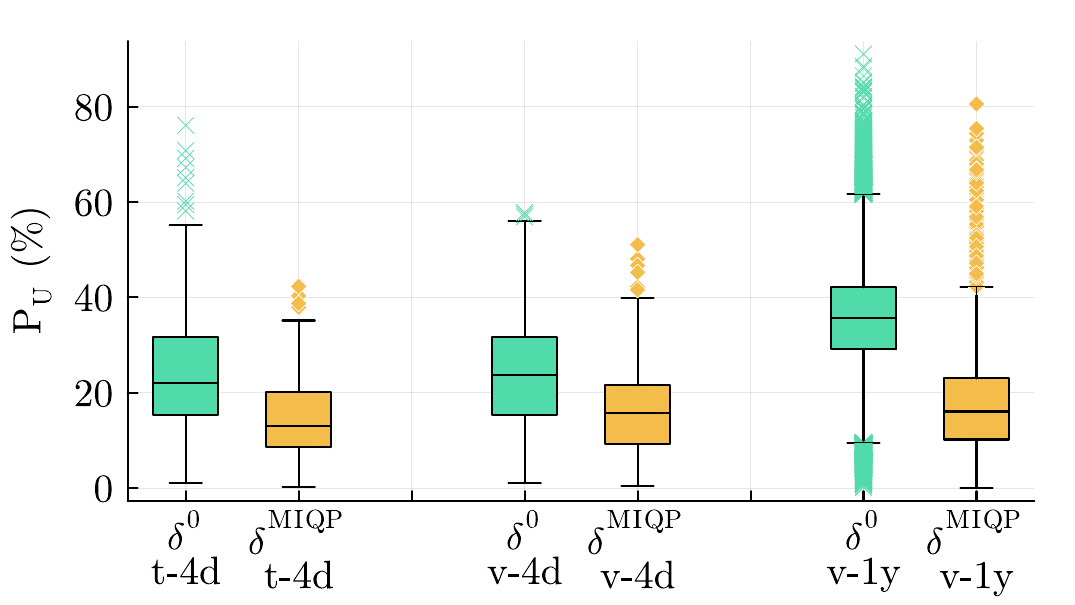}\label{fig:b}}
    
    \caption{Distributions of PVUR (left) and P$_{\text{U}}$ (right) imbalance per time step for the \textit{training} load-data, i.e. the four days under optimization (t-4d), and \textit{validation} load-data, namely the subsequent four days (v-4d), and the entire year (v-1y). The results for default configuration ($\bm{\delta^0}$) and the MIQP solutions ($\bm{\delta^{\text{MIQP}}}$) are shown in green and yellow respectively.}
    \label{fig:validation_MIQP_sols}
    \vspace{-3mm}
\end{figure*}

\vspace{-2mm}

\subsection{Solution validation for unseen loads (TC1)}

Figure \ref{fig:validation_MIQP_sols} presents the distributions of imbalance PVUR$_t$ and P$_{\text{U},t}$ per time step $t \in \timeseriesm$, i.e. the first 4 days of January (training-4d). These results are shown for both the default configuration ($\bm{\delta^0}$) and the MIQP solutions from \textit{TC1} ($\bm{\delta^{\text{MIQP}}}$) for PVUR$^*$ (left figure) and P$_\text{U}^*$ (right figure). Additionally, to assess whether the imbalance reduction is maintained for unseen loads, the distributions are presented for the subsequent 4 days (validation-4d), and for the entire year (validation-1y).

In all cases, the MIQP solution has lower and less variable imbalance than the default configuration. Hence the solution found for the first 4 days of January already greatly improves the situation for the entire year. Nevertheless, since only the average imbalance is optimized, outliers can still occur at specific timesteps. For example, in the PVUR$_t$ distribution (with $t~\in~\text{t-4d}$), the worst imbalance outlier happens for the MIQP solution $\bm{\delta^{\text{MIQP}}}$ rather than for the default configuration $\bm{\delta^0}$, despite $\bm{\delta^{\text{MIQP}}}$'s overall better performance. Potential remedies to mitigate these outliers is to enforce a hard constraint on the maximum imbalance. If this proves infeasible, phase reconfiguration could be complemented with other balancing strategies, such as demand-response programs to shift peak imbalances.

\vspace{-4mm}

\subsection{Impact of the number of switching actions (TC2)}

\begin{figure*}[t!]
    \centering
    {\includegraphics[width=0.475\textwidth]{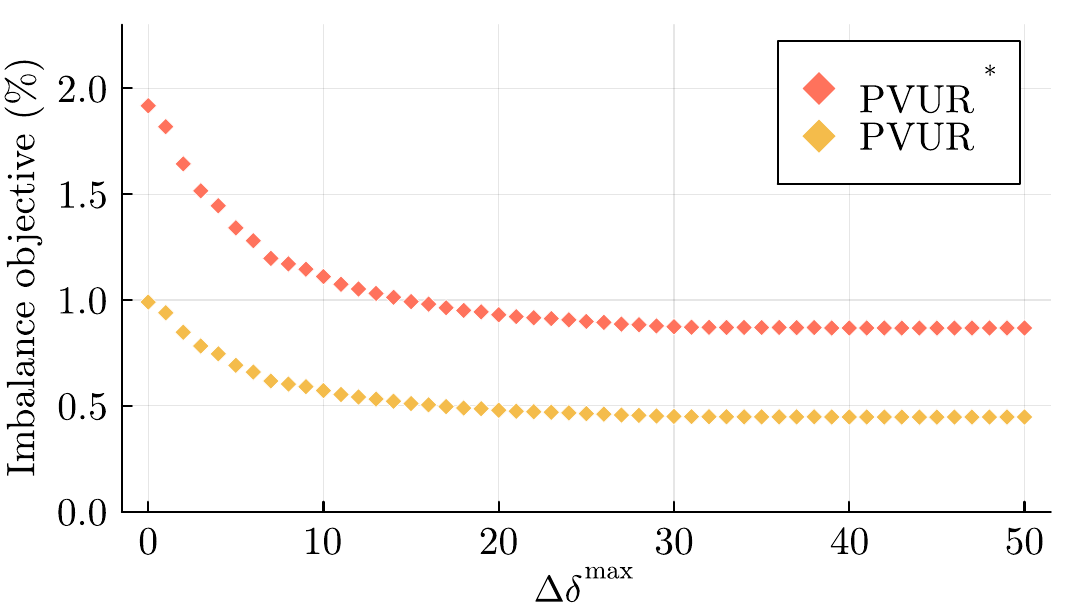}\label{fig:a}}
    \qquad
    {\includegraphics[width=0.475\textwidth]{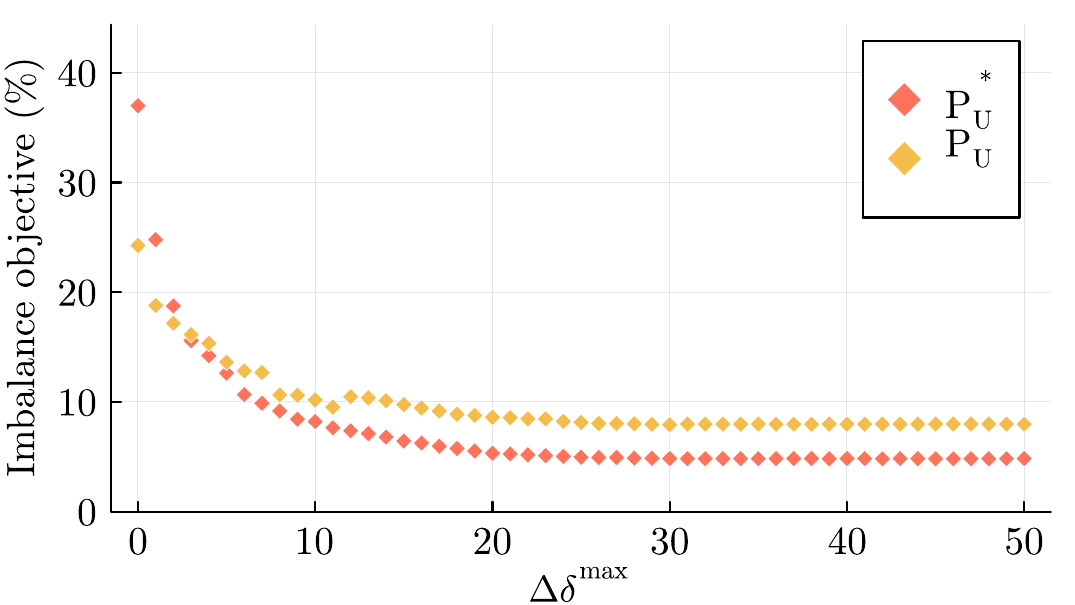}\label{fig:b}}
    
    \caption{Impact of the maximum allowed number of switching actions $\Delta{\delta}^{\text{max}}$ on the imbalance objectives PVUR, and PVUR$^*$ (left), P$_\text{U}$, and P$_\text{U}^*$ (right). Each optimization was done for 24 time steps (single day with hourly resolution).}
    \vspace{-5mm}
    \label{fig:lim_SA_24ts}
\end{figure*}

Figure \ref{fig:lim_SA_24ts} shows the results of \textit{TC2}, illustrating the impact of the maximum allowed number of switching actions $\Delta{\delta}^{\text{max}}$ on the imbalance objectives.
Unsurprisingly, higher $\Delta{\delta}^{\text{max}}$ allows for more imbalance reduction. For all objectives, the imbalance decrease seems exponentional in function of $\Delta{\delta}^{\text{max}}$ with a saturation around $\Delta{\delta}^{\text{max}}=30$, that is 55\% of the total number of consumers $|\singlephaseusersetm|$ of the given feeder. Setting $\Delta{\delta}^{\text{max}}$ to 5, i.e. 10\% of $|\singlephaseusersetm|$, already results in 50\% of the total achievable imbalance decrease illustrating the trade-off operators have to make between limiting the time and money required for the maintenance action, while obtaining sufficient imbalance reduction.

\subsection{Scalability of the methods (TC3)}

\begin{figure}[t]
\centering
\includegraphics[width=0.5\linewidth]
{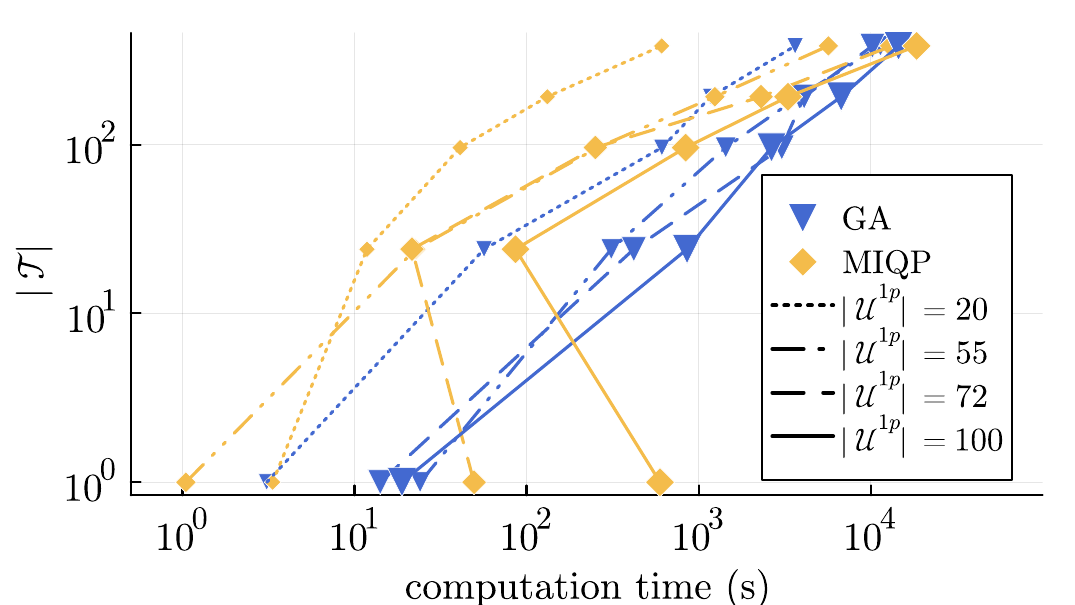}
\caption{Relation between the optimization horizon (number of time steps considered $|\timeseriesm|$) and the total computation time. MIQP and GA results are shown in yellow and blue respectively. Four different feeders with 20, 55, 72, and 100 consumers are represented with increasing marker size. }
\label{fig:Scalability}
\end{figure}

The results of the scalability test case (\textit{TC3}) are presented in Figure \ref{fig:Scalability},  illustrating the optimization horizon (number of time steps considered $|\timeseriesm|$) in function of the computation time for both MIQP (yellow diamond marks) and GA (blue triangle marks) and for four different feeders with 20, 55, 72, and 100 consumers $|\singlephaseusersetm|$. The MINLP did not converge within 24 hours for these feeders. For a small feeder of four consumers, over a two-day, 15-minute interval scenario, MINLP convergence required six hours.

On the log-log scale, the GA method appears to exhibit a linear trend, indicating an exponential relationship between the number of time steps ($|\timeseriesm|$) and computation time. The MIQP method shows a similar linear trend, with the exception of the single time-step case ($|\timeseriesm| = 1$), which deviates from the pattern by requiring disproportionately more computation time. This anomaly could be due to the presence of solution symmetries in the single-step scenario resulting in many similar solutions causing slower convergence.

The GA method shows a steeper slope than MIQP, implying that GA may scale better over longer time horizons, which is further supported by MIQP becoming slower than GA for some of the larger test cases. In general, the computation time tends to increase significantly with larger $|\singlephaseusersetm|$. However, this relationship is less consistent than the exponential trend observed for $|\timeseriesm|$, for example feeders with more consumers sometimes require less or similar computation time than feeders with less consumers. Still, European 
LV feeders are typically designed for around 40 consumers  (with a 250~~kVA~MV$/$LV transformer rating), and while networks with 100 customers exist, they are rare. Thus, we can reasonably assume that mathematical programming is best suited for these applications. 

\section{Conclusion}\label{sec:conclusion}
This study develops and compares several calculation methods to reduce phase imbalance in LV distribution grids through \textit{static} phase reconfiguration, often more practical and  cost-effective than its \textit{dynamic} counterpart. Three methods are presented: an exact MINLP formulation, an MIQP approximation using LinDist3Flow, and a genetic algorithm (GA). These methods are more expressive than previous static models and support key imbalance metrics, phase voltage unbalance ratio (PVUR) and power unbalance (P$_U$), with proxy objectives (PVUR$^*$ and P$_U^*$) for MIQP.

The three methods were evaluated on a 55-consumer ENWL feeder over four days. The results show that using expressive models enables operators to select the most relevant imbalance metric; PVUR reduces voltage imbalance (up to 27\% reduction), and P$_U$ targets power imbalance (up to 40\%). The methods were compared for computational performance: the MINLP failed to converge within 24 hours, while the MIQP, despite using proxy objectives, consistently outperformed the GA. The MIQP achieved 7\% and 10\% more imbalance reduction than the average GA result for PVUR and  P$_U$ respectively. Despite challenging hyperparameter tuning, the GA method suffered from high result variability and had slower computation times (110 minutes, compared to the MIQP's 50 minutes for PVUR and 90 minutes for P$_U$).

Reconfigured feeders maintained lower imbalance levels and reduced imbalance variation even for a validation period of up to one year. Additionally, analysis demonstrated that switching only 10\% of consumers could achieve 50\% of the total imbalance reduction, highlighting a trade-off between maintenance cost and performance. Scalability tests showed that the methods’ computational time increased exponentially with larger feeders and longer time horizons; MIQP remained faster for most cases but seemed to show worse scalability for the larger test cases.

Overall, the results emphasize the advantage of using expressive models in static phase reconfiguration. The MIQP approach, despite using proxy objectives, provides a robust and efficient solution for mitigating phase imbalance in LV grids, offering a viable method for grid operators in real-world applications.

\section*{Acknowledgment}
This work received funding from the Research Foundation Flanders (FWO) through Grant 1SA7222N for strategic basic research.

\bibliographystyle{IEEEtran}
\bibliography{IEEEabrv,Bibliography_abb}

\end{document}